\documentclass[aps,prl,twocolumn,showpacs,superscriptaddress,groupedaddress]{revtex4}  
\usepackage{subfigure}
\usepackage{graphicx}  
\usepackage{dcolumn}   
\usepackage{bm}        
\usepackage{amssymb}   
\usepackage{mathtools}
\usepackage{color}
\usepackage{slashed}
\usepackage{soul}
\usepackage{times}
\usepackage{xspace}

\def\jraa{R_{\tiny{\rm AA}}}

\def\aSC{{\kappa_{\rm sc}}}

\begin{document}



\title{Jet suppression from small to intermediate to large radius}
\date{\today}

\author{Daniel Pablos}
\affiliation{Institutt for fysikk og teknologi, University of Bergen, Postboks 7803, 5020 Bergen, Norway}

\begin{abstract}

We present predictions for jet suppression from small to intermediate to very large radius, for low and very high energy jets created in heavy ion collisions at the LHC. We use the hybrid strong/weak coupling model for jet quenching that combines perturbative shower evolution with an effective strongly coupled description of the energy and momentum transfer from the jet into the hydrodynamic quark-gluon plasma. Because of momentum conservation, the wake created by the jet enhances or depletes the amount of particles generated at the freeze-out hypersurface depending on their orientation with respect to the jet. Within such framework we find that jet suppression is surprisingly independent of the anti-$k_T$ radius $R$, first slightly increasing as one increases $R$, then at larger values of $R$ very slowly decreasing.
This nearly independence of jet suppression with increasing values of $R$ arises from two competing effects, namely the larger energy loss of the hard jet components, which tends to increase suppression, versus the partial recovery of the lost energy due to medium response, reducing suppression. We also find that the boosted medium from the recoiling jet reduces the amount of plasma in the direction opposite to it in the transverse plane, increasing the amount of jet suppression due to an over-subtraction effect. We show that this characteristic signature of the hydrodynamization of part of the jet energy can be quantified by selecting samples of dijet configurations with different relative pseudorapidities between the leading and the subleading jet.


\end{abstract}


\maketitle

\textit{Introduction.}
The strong yield suppression and substructure modification observed in the analysis of the high $p_T$ jets produced in heavy ion collisions compared to those measured in nucleon-nucleon collisions is ascribable to the production of extended, deconfined QCD matter. These modifications, typically referred to as jet quenching phenomena \cite{Mehtar-Tani:2013pia}, arise from the interaction of the energetic colored charges formed through parton showers with the QCD medium. The strong correlations among the thousands of low $p_T$ particles created in such nucleus-nucleus collisions can be very well described by hydrodynamic simulations of an exploding droplet of hot QCD liquid  \cite{Acharya:2018zuq}, known as the quark-gluon plasma (QGP). These simulations are surprisingly successful at describing the comparable in magnitude flow-like signals observed in smaller systems such as nucleon-nucleus and nucleon-nucleon collisions at high multiplicity \cite{Weller:2017tsr}.


 One of the key aspects necessary to assess the fluid nature of the QGP consists in the determination of the measurable consequences of the response that the flowing medium would have under the passage of an energetic jet. After crossing an infinite bath of deconfined plasma, such probe would necessarily have to become part indistinguishable of the fluid through a process of energy degradation - it would have to \emph{hydrodynamize}. The rate at which such hydrodynamization occurs depends strongly on the theoretical framework used to describe the process of energy loss, which can yield wildly different answers for the spatial distribution of the total energy of the jet after quenching. 
On general grounds, perturbative approaches where the coupling with the medium is required to be small 
 typically need a rather long in-medium path in order to hydrodynamize a significative fraction of the original energy of the jet. 
 Complementarily, non-perturbative strongly coupled computations present, as it is natural, a much faster hydrodynamization rate due to a very small mean free path.
 Being able to experimentally access the process of energy hydrodynamization represents in this way not only an invaluable channel with which to understand the emergence of collectivity in QCD 
 but also allows us to confront the very different pictures of the inner workings of the QGP 
 under which such hydrodynamization process is described.
 
 In this work we study the observable effects of energy and momentum hydrodynamization in the phenomenological description of jet suppression in heavy ion collisions. Jets are extended objects consisting of a collection of hadrons that are clustered together after choosing a specific reconstruction algorithm with a given jet radius parameter $R$. 
 The larger the radius $R$, the more extended in $(\eta,\phi)$ space the jet can be, which in particular means that the jet is wider and will typically be the result of the fragmentation of a larger number of partons. Due to momentum conservation, the energy that those partons deposit in the fluid is correlated with the jet direction, and as such is part of the jet signal that is measured in experiments after the uncorrelated background is subtracted.
 
Given that wide jets containing a larger number of sources of energy loss will tend to lose more energy than the narrower ones, based on this consideration alone the amount of jet suppression should increase with increasing jet radius $R$. However, having a larger jet radius $R$ also means that the jet will retain a larger fraction of the widely distributed energy and momentum deposited in the plasma. These competing effects yield a jet suppression that is remarkably independent of the jet radius $R$ across a vast range in jet $p_T$ . We further show that the medium response caused by the energy deposited by the recoiling jet has sizeable effects on the signal measured by the trigger jet. These striking long-range correlations, which depend on the pseudorapidity separation between the dijet system, are a salient feature of the assumed hydrodynamization of a large fraction of the energy of a jet and as such represent a distinctive signature of the fluidlike behaviour of the QGP.

\textit{The hybrid strong/weak coupling model.} High energy jet production and evolution has been very successfully described in vacuum through perturbative QCD, both analytically and numerically through the use of Monte Carlo event generators. Jets created in heavy ion collisions are known to experience important modifications with respect to those created in proton-proton collisions, which relate both to initial state effects due to the nuclear modification of the nucleons parton distribution functions (PDFs) as well as the more dramatic final state effects associated to the interaction of the parton shower with the QGP. Even though the initial evolution of an energetic jet will be dominated by the high virtuality associated to its production, $Q^2 \sim \mathcal{O}(p_T^2)$, the presence of the medium introduces a new scale in the problem, its temperature $T \ll Q^2$, which is of the order of the non-perturbative QCD scale $T \sim \mathcal{O}(\Lambda_{\rm QCD})$. By assuming that this non-perturbative scale dominates the physics of the interaction between an energetic probe and the QGP, it is evident that perturbative techniques alone might not suffice to describe all the dynamics behind jet quenching phenomena. 

The wide scale separation allows nevertheless for an effective description, the hybrid strong/weak coupling model used in this work \cite{Casalderrey-Solana:2014bpa,Casalderrey-Solana:2015vaa,Casalderrey-Solana:2016jvj}, in which the high virtuality relaxation process through successive splittings is described perturbatively, at weak coupling, and the non-perturbative interaction of each of the partons with the QGP is described at strong coupling by the use of holographic techniques. Within this model we use the event generator PYTHIA \cite{Sjostrand:2014zea}, supplemented with the LO nuclear modifications to the PDFs as calculated in \cite{Eskola:2009uj}, to create and evolve high $p_T$ parton showers, to which we assign a space-time description through a formation time argument such that each parton takes a time $\tau_f = 2 E /Q^2$ to split. In between splittings, partons transfer energy and momentum to the plasma hydrodynamic modes according to a strongly coupled energy loss rate that was derived within holography for $\mathcal{N}=4$ SYM, at large $N_c$ and infinite coupling \cite{Chesler:2014jva,Chesler:2015nqz}, which reads

\begin{equation}\label{CR_rate}
\left. \frac{d E_{}}{dx}\right|_{\rm strongly~coupled}= - \frac{4}{\pi} E_{\rm in} \frac{x^2}{x_{\rm therm}^2} \frac{1}{\sqrt{x_{\rm therm}^2-x^2}} \quad , 
\end{equation}

where $E_{\rm in}$ is the parton's initial energy and $x_{\rm \small therm}=(E_{\rm in}^{1/3}/T^{4/3})/2 \aSC$ is the maximum length that the parton can travel within the plasma before completely hydrodynamizing, with $T$ being the local temperature of the plasma. An important modelling assumption is that all differences in the energy loss experienced by a parton in a QCD plasma compared to a $\mathcal{N}=4$ SYM plasma can be encapsulated in the 't Hooft coupling dependent parameter $\aSC$ \cite{Casalderrey-Solana:2014bpa}, which we fixed through a global comparison to hadron and jet data at the LHC \cite{Casalderrey-Solana:2018wrw}. The local properties of the QGP needed to apply the energy loss formula in Eq.~\ref{CR_rate} are read from event-averaged hydrodynamic profiles that successfully reproduce the measured spectra and flow features of low $p_T$ hadrons \cite{Shen:2014vra}. Partons that are not completely hydrodynamized are fragmented into hadrons using the Lund string model included in PYTHIA.

It has been shown that the deposition of energy and momentum from a localized source in a strongly coupled plasma generates an hydrodynamic wake after very short time scales, of $\mathcal{O}(1/T)$ \cite{Chesler:2007an}. In order to estimate the measurable effects of such modifications of the QGP we can compute how the jet induced perturbations in the energy-momentum tensor of an ideal, boost-invariant fluid translate into a modification of the final hadron distribution associated with the `particlization' process that occurs after the plasma cools down at a certain hypersurface \cite{Casalderrey-Solana:2016jvj}. After assuming that the induced perturbations in the four-velocity and the entropy of the fluid are small, and by noting that jets propagate through the plasma approximately at a fixed space-time rapidity, we can use the standard Cooper-Frye prescription \cite{Cooper:1974mv} to express the distribution of the hadrons coming from the wake \cite{Casalderrey-Solana:2016jvj}, as 

\begin{equation}
\label{onebody}
\begin{split}
E & \frac{d\Delta N}{d^3p}=\frac{1}{32 \pi} \, \frac{m_T}{T^5} \, \cosh(y-y_j)  \exp\left[-\frac{m_T}{T}\cosh(y-y_j)\right] \\
 &\times \Bigg\{ p_{T} \Delta P_{T} \cos (\phi-\phi_j) +\frac{1}{3}m_T \, \Delta M_T \, \cosh(y-y_j) \Bigg\} \quad ,
\end{split}
\end{equation}

\begin{figure*}[t!]
\includegraphics[width=1\textwidth
]{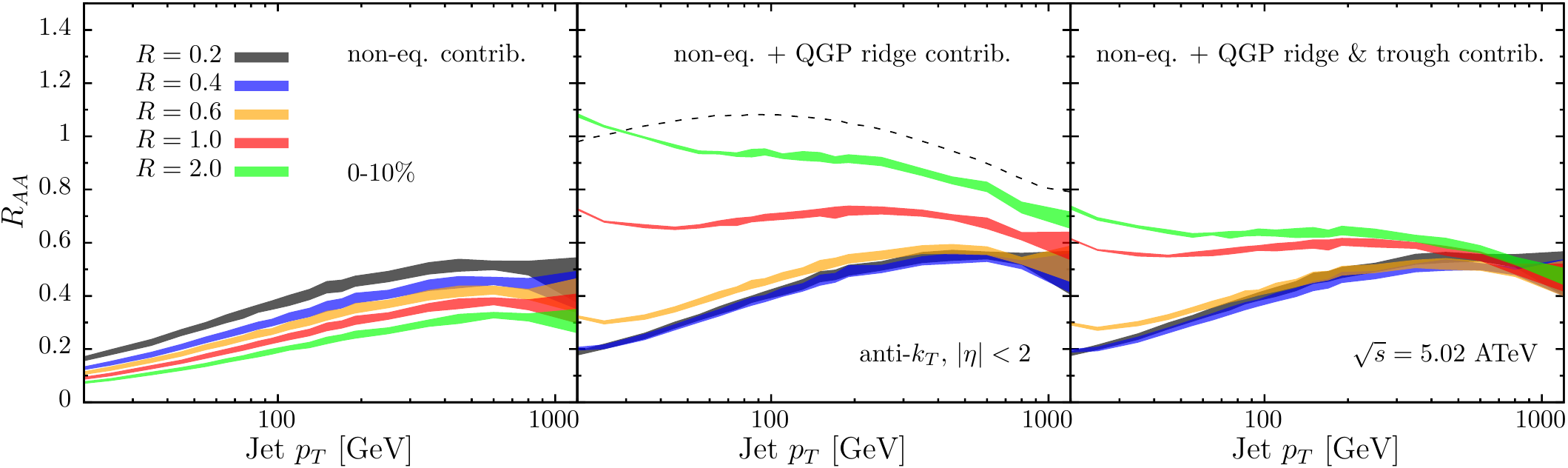}
\vspace{-0.2in}
\caption{\label{fig:raavsr} The anti-$k_T$ jet radius $R$ dependence of $\jraa$ in PbPb collisions at $\sqrt{s}=5.02$ ATeV for the 0-10\% centrality class. The left (middle) panel contains only the non-equilibrium (non-equilibrium + QGP ridge) contribution, while the full result that includes the effects of the QGP trough is shown in the right panel. The dashed black line in the middle panel shows the result for $\jraa$ of $R=2$ jets without quenching.}
\end{figure*}

where $T$ is the temperature at the hypersurface, $m_T \equiv \sqrt{p_{T}^2 +m^2}$, $y$ and $\phi$ are the transverse energy, rapidity and azimuthal angle of the hadron, respectively, while $\phi_j$ and $y_j$ are the azimuthal angle and rapidity of the parton that deposited an amount of transverse momentum $\Delta P_{T}$ and transverse energy $\Delta M_T$ in the plasma. This distribution can become negative, most notably along the direction opposite to the propagation of the wake. This simply means that in such region there is a depletion of the flowing medium with respect to the unperturbed QGP due to the boost experienced by the fluid \footnote{This depletion effect has also been reported in analysis where the deposited energy is treated as a source term in the hydrodynamic evolution equations \cite{Chen:2017zte}.}.
Such depletion of the plasma, which is an inseparable part of the jet signal, can amount to energy loss, or jet suppression, due to an over-subtraction effect after the uncorrelated background is subtracted as it is typically done in experiments.
A useful way to picture this effect is by imagining the shape of a wave on the surface of a fluid: the \emph{ridge} of the wake would correspond to the positive contributions from Eq.~\ref{onebody}, while the \emph{trough} would be associated to the negative ones. For each of the partons that interacts with the plasma we sample this closed distribution and generate hadrons whose collective energy and momentum exactly match the energy and momentum deposited by the parton.

In this way, the initial four-momentum in the event has been redistributed into hadrons with distinct origins: the ones coming from the hadronization of well identified partons, which we can call the non-equilibrium contribution, and those coming both from the excess and depletion of the fluid QGP, which we can call the QGP ridge and QGP trough contributions, respectively.

\textit{Jet suppression as a function of the opening angle.} The results in Fig.~\ref{fig:raavsr} show the yield suppression of jets created in PbPb collisions at $\sqrt{s}=5.02$ ATeV for the 0-10\% centrality class compared to those in pp collisions at the same energy, quantified as it is customary in experiments through the variable $\jraa(p_T) \equiv (dN^{\rm AA}/dp_T)/(dN^{\rm pp}/dp_T)/\langle N_{\rm coll} \rangle$, which is normalized to the average number of collisions in PbPb for the given centrality class, $\langle N_{\rm coll} \rangle$, for different values of the radius $R$ using the anti-$k_T$ reconstruction algorithm encoded in FastJet \cite{Cacciari:2011ma}. In order to best illustrate how the effects of the medium response affect this observable, along with the full results shown in the right panel of Fig.~\ref{fig:raavsr} we provide two clarifying, yet unphysical scenarios in the left and middle panels. The results shown in the left panel were obtained by taking into account only the hadrons generated through the fragmentation of the non-hydrodynamized jet partons, the non-equilibrium contribution. The visible trend of increasing suppression with increasing $R$ is due to the fact that by opening the jet cone, for a given jet $p_T$ one can select jets with a softer and wider fragmentation, increasing the number of energy loss sources and the total amount of energy transferred to the QGP. The fact that wider structures are relatively more suppressed than the narrower, less active ones, is the reason of the success of the present model to describe recent experimental results from ALICE regarding jet substructure modification using grooming techniques \cite{Acharya:2019djg}. While we showed that the effects from the wake are negligible for these kind of observables due to the robustness of the grooming procedure to the presence of soft particles \cite{Casalderrey-Solana:2019ubu}, the contribution from the hydrodynamized jet energy plays a leading role in the description of jet suppression as we will see next.

The panel in the middle of Fig.~\ref{fig:raavsr} contains both the non-equilibrium particles and the excess of hadrons coming from the decay of the hydrodynamic wake generated by the passage of the jet, the QGP ridge contribution. The inclusion of these soft hadrons, spread over a wide distribution in the azimuthal angle $\phi$ with respect to the jet direction, reduces jet suppression more the larger the jet radius $R$ is. While jets with a radius of $R=0.2$ barely capture any of these hadrons, jets with $R=0.4$, in contrast to what is seen in the left panel, are now equally suppressed as those with $R=0.2$ even though the total amount of energy transferred to the plasma is larger. Given that jets typically are collimated structures, total energy loss tends to saturate as a function of the radius $R$, which means that after some intermediate value it is natural to expect that the jet retains a larger fraction of the initial hard parton momentum than compared to any smaller value of $R$. From a radius of $R=0.4$ onwards, jet suppression decreases slowly, but steadily, being necessary to go up to $R\sim 2$ to get back most of the lost energy, or equivalently to values of $\jraa \lesssim 1$. Even though the energy is practically fully recovered within very large cones, $\jraa$ visibly decreases at high momentum, starting from $p_T \gtrsim 300$ GeV. This sizeable effect is largely due to the modification of the initial jet spectrum in PbPb induced by the nuclear effects on the PDFs at large values of $x$, as can be seen by the dashed black line in the middle panel of Fig.~\ref{fig:raavsr} where we show $\jraa$ for $R=2$ jets without quenching. 
The sensitivity of jet suppression to such initial state effects is interesting per se, and can be used in the near future as a new channel with which to constrain nuclear PDFs through the standard global fit procedures.

We now discuss the right panel of Fig.~\ref{fig:raavsr}, where all hadrons are included in the analysis. This is the actual physical situation in which the non-equilibrium contributions together with the hydrodynamized QGP ridge and trough are balanced such that the four-momentum in the event is conserved. Comparing with the middle panel of Fig.~\ref{fig:raavsr}, where only the QGP ridge contribution was included, we can observe that jet suppression has been increased, specially for the larger values of $R$. The origin of this suppression lies in the depletion of the fluid around the jet axis induced by the passage of the \emph{recoiling jet}. Whenever the cone of a given jet captures the contribution from the QGP trough induced by the fluid modification of the back-to-back jet, after subtracting what would correspond to the uncorrelated background, e.g. the background density estimated from minimum-bias events of the same centrality class where no high $p_T$ jet is present, the energy of the jet is reduced through an over-subtraction effect. All in all, the contribution from the non-equilibrium hadrons and those coming from the QGP ridge and trough result into a jet yield suppression that is remarkably insensitive to the radius parameter $R$ at high $p_T$. We predict that in order to observe a sizeable $R$ dependence one would need to go to fairly large angles, $R\gtrsim 1$, and relatively low momentum, $p_T\lesssim 200$ GeV, a combination of values which is particularly hard to access experimentally due to the reduction of the signal to background ratio for jets with large cones at small momentum. There is, nevertheless, a way in which the effect of the QGP trough can be quantified at relatively high jet $p_T$ by exploiting the lack of rapidity correlation in dijet pairs created in hadronic collisions. 

\begin{figure}
\hfill
\includegraphics[width=0.4\textwidth
]{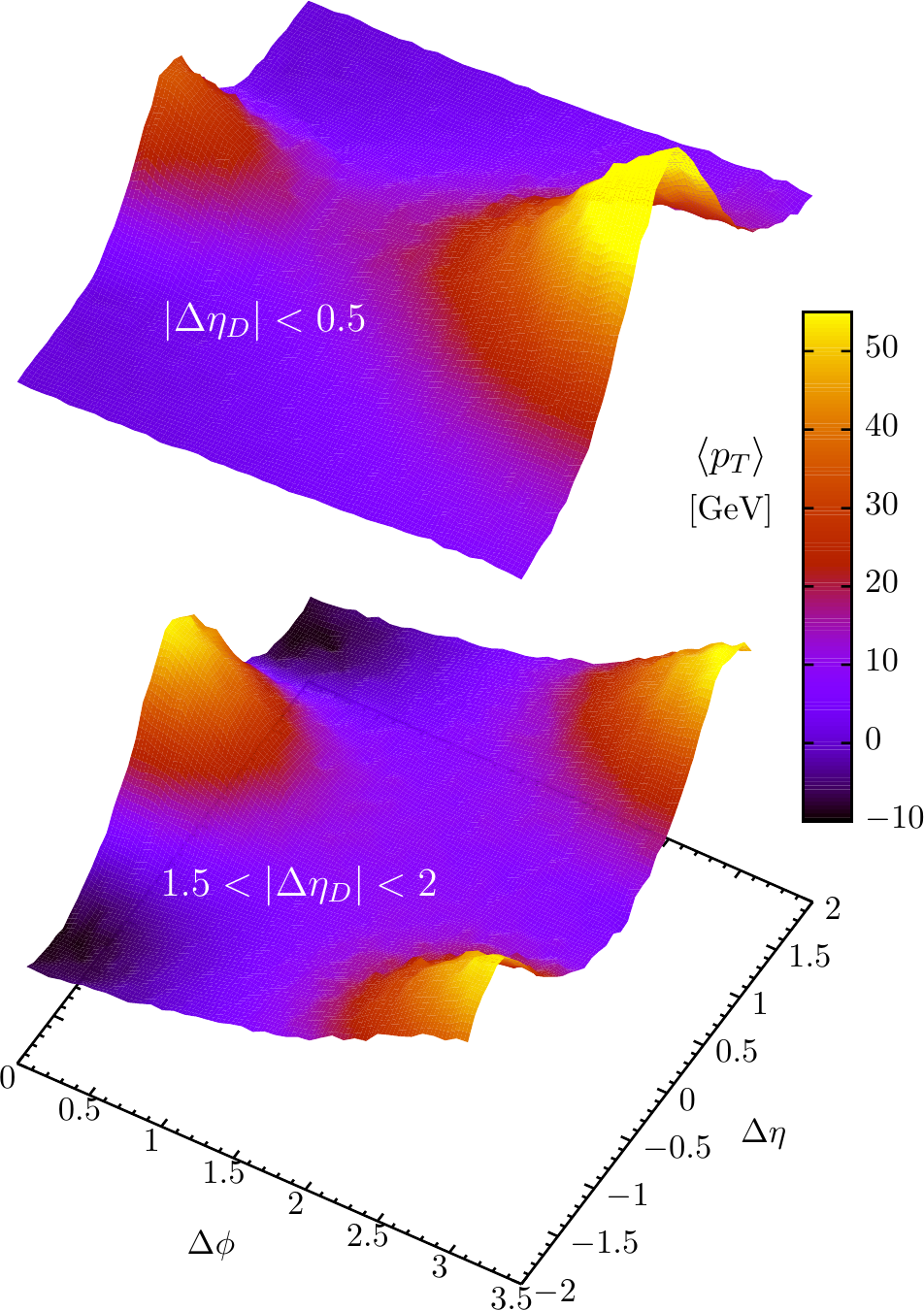}
\vspace{-0.1in}
\caption{\label{fig:wakedens} The $\langle p_T \rangle$ density of the particles coming from the wake with respect to the leading jet direction, in terms of $\Delta \eta$ and $\Delta \phi$. The leading jet has $p_T^L>250$ GeV and the subleading has $p_T^S>80$ GeV, with an angular separation in the transverse plane $\Delta \phi_D > 2\pi/3$ and an absolute difference in rapidity $|\Delta\eta_D|$.}
\end{figure}

\textit{The effect of the recoiling jet.} The center of mass of the hard partonic collision that produces a dijet system is in general different from the laboratory frame in which observables are measured. This obvious fact, together with the comparatively much narrower $\Delta \eta$ distribution of the soft hadrons coming from the wake with respect to their sourcing jet, as described by Eq.~\ref{onebody}, allows us to select samples of dijets where we can engineer the amount of QGP trough contribution that the leading jet captures within its cone of radius $R$. To this end, we choose to study dijet pairs with $|\eta|<2$ where the leading jet has $p_T^L>250$ GeV and the subleading has $p_T^S>80$ GeV, with an angular separation in the transverse plane $\Delta \phi_D > 2\pi/3$ and an absolute difference in pseudorapidity $|\Delta\eta_D|$, which we vary. A very convenient way to visualize the contributions from the QGP ridge and trough in a dijet system is by plotting the $\langle p_T \rangle$ density of the hydrodynamized particles as a function of their separation in azimuthal angle $\Delta \phi$ and pseudorapidity $\Delta \eta$ with respect to the leading jet, as we show in Fig.~\ref{fig:wakedens}. The top plot in Fig.~\ref{fig:wakedens} corresponds to dijet configurations of jets with $R=1$ which are close in pseudorapidity, namely $|\Delta\eta_D|<0.5$, while the dijet samples from the bottom plot have a larger separation of $1.5<|\Delta\eta_D|<2$. In the top plot we can clearly see the QGP ridge contribution both at the near side, with $\Delta \phi \sim 0$ and at the away side, with $\Delta \phi \sim \pi$. The size of the QGP ridge is greater in the away side because the subleading jet has on average lost more energy to the plasma than the leading jet at the near side due to its lower jet $p_T$ cut. The negative $\langle p_T \rangle$ density contribution associated to the QGP trough generated by the subleading jet cannot be seen with clarity in the top plot because it falls on top the QGP ridge contribution generated by the wake of the leading jet. When the dijet system is well separated in pseudorapidity, as presented in the lower plot of Fig.~\ref{fig:wakedens}, we can see a black dip of depleted fluid at the near side, centered around the pseudorapidity at which the subleading jet sits with respect to the leading jet. The fact that there is barely any signal of the QGP trough in the away side basically reflects that the leading jet has on average lost comparatively much less energy than the subleading jet. What this means in terms of jet energy loss is that in the situations in which the QGP trough generated by the subleading jet does not hit the area defined by the leading jet cone of radius $R$, the latter experiences less suppression and its $\jraa$ is increased compared to the case in which the pseudorapidity separation is small.

\begin{figure}
\includegraphics[width=0.4\textwidth
]{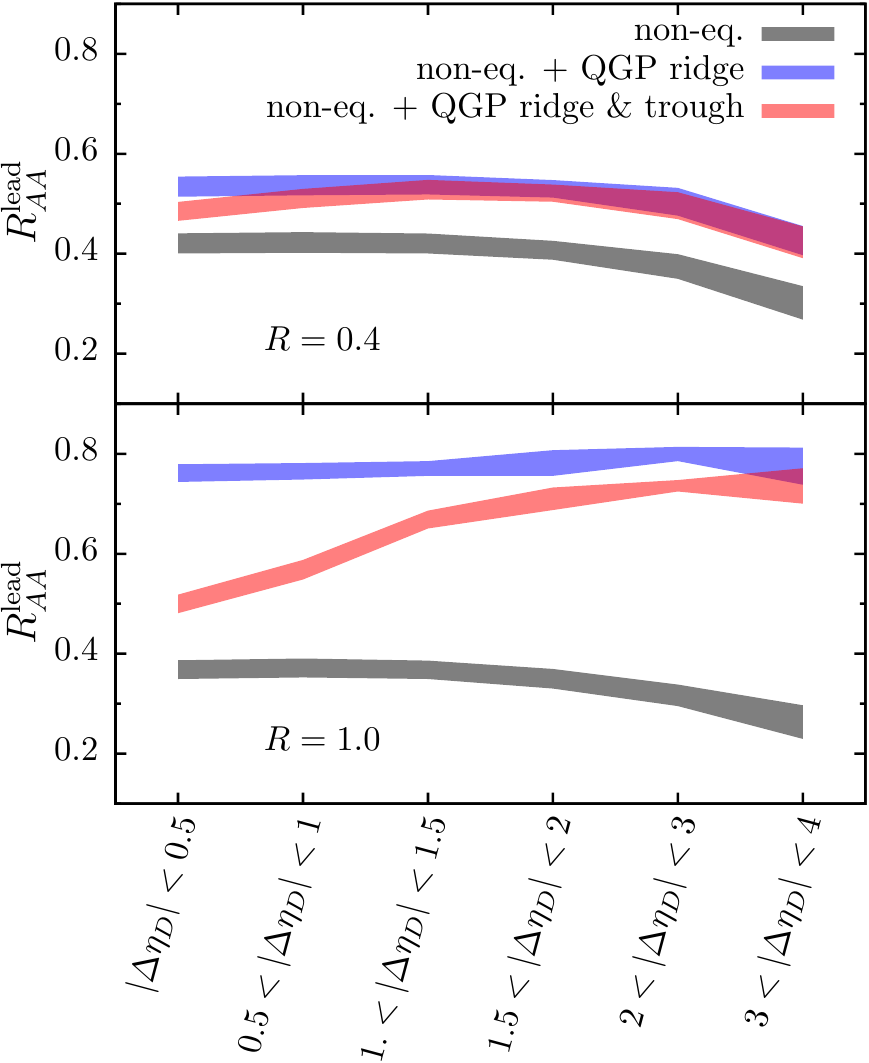}
\vspace{-0.1in}
\caption{\label{fig:deltaeta}  Jet suppression $R_{AA}^{\rm lead}$ for leading jets above $p_T^L>250$ GeV, as a function of the pseudorapidity separation $|\Delta\eta_D|$ with respect to the back-to-back subleading jet with $p_T^S>80$ GeV. Top panel is for jets with an anti-$k_T$ radius $R=0.4$, while lower panel is for jets with $R=1$.}\end{figure}

This dependence on jet suppression as a function of the pseudorapidity separation between a dijet system can be easily quantified by computing the leading jet yield suppression, which we can call $R_{AA}^{\rm lead}$, as a function of the $|\Delta\eta_D|$ between the leading and subleading jets. In Fig.~\ref{fig:deltaeta} we present such computation for intermediate jet radius with $R=0.4$ in the top panel and large jet radius $R=1$ in the bottom panel. Similarly to Fig.~\ref{fig:raavsr}, we present three sets of results corresponding to the non-equilibrium contribution only, in black, the non-equilibrium plus the QGP ridge, in blue, and all the contributions, including the QGP trough, in red. The first thing we observe is the approximate flatness of leading jet suppression for a wide range of $|\Delta\eta_D|$ for the two sets of results that omit the contribution from the QGP trough, both for $R=0.4$ and $R=1$ jets. The reason of the drop of $R_{AA}^{\rm lead}$ at the highest values of $|\Delta\eta_D|$ is because of the hard parton $p_T$ spectrum becoming steeper due to the requirement that both the leading and subleading jets sit at relatively large pseudorapidities. 
The dependence on $|\Delta\eta_D|$ comes when we include the effect of the QGP trough generated by the wake of the more quenched subleading jet. As expected from simple geometrical considerations, the effect of the recoiling jet is gradually reduced as one increases $|\Delta\eta_D|$, being almost irrelevant beyond $|\Delta\eta_D| \gtrsim R + \sigma(\Delta \eta)$, where $\sigma(\Delta \eta)\simeq 1$ symbolizes the range in pseudorapidity below which most of the particles from the wake sit, as described by Eq.~\ref{onebody}. Interestingly, given that most of the hydrodynamized energy is sourced from the centre of the collimated jets,
the rate at which $R_{AA}^{\rm lead}$ increases with $|\Delta\eta_D|$ presents a knee around $|\Delta\eta_D|\sim R$, as can be more clearly appreciated in the bottom panel of Fig.~\ref{fig:deltaeta}.

\textit{Conclusions.} The  angular dependence of jet suppression encodes key information about the process of energy and momentum hydrodynamization, and for this reason can be used to greatly improve our understanding of fundamental aspects of the jet/QGP interaction. Distinctive features of the efficient hydrodynamization process that follows our assumption of energy loss at strong coupling are seen through the remarkable independence of inclusive jet suppression from the jet opening angle $R$, which in particular means that $\jraa$ does not go to $1$ even for very large radius, contrary to current expectations.  We have shown that this fact does not contradict energy and momentum conservation and that it is a consequence of the effect of the trough associated to the wake generated by the recoiling jet. The dependence of leading jet suppression on the pseudorapidity separation of a dijet system allows us to present a set of predictions for the discovery of the effect of the QGP trough and, by extension, one of the most significant tests  from jet quenching dynamics of the fluidlike behaviour of the QGP.

\textit{Acknowledgments.} We thank Jorge Casalderrey-Solana, Yi-Lun Du, Chris McGinn, Krishna Rajagopal, Yen-Jie Lee, \'Ad\'am Tak\'acs and Konrad Tywoniuk for useful discussions.
This work is supported by a grant from the Trond Mohn Foundation (project no. BFS2018REK01).


\end{document}